\documentclass{llncs}
\usepackage{amssymb,amsmath,latexsym}
\usepackage{graphicx}        
\usepackage{verbatim}
\usepackage{multicol}
\usepackage{hyperref}
\usepackage{bigfoot}

\title{A general translation from nested Petri nets into PROMELA\thanks{This work is an extended version of the one presented in~\cite{venero2013a}. It was supported by the S\~ao Paulo Research Foundation (FAPESP) under the grant 2010/52505-0.}}

\author{{Mirtha Lina Fern\'andez Venero} \and {Fl\'avio Soares Corr\^ea da Silva} }
\institute{Department of Computer Science, University of S\~ao Paulo, Brazil, 05508-090 \email{\{mirtha, fcs\}@ime.usp.br} }

\newenvironment{keywords}{
       \list{}{\advance\topsep by0.35cm\relax\small
       \leftmargin=1cm
       \labelwidth=0.35cm
       \listparindent=0.35cm
       \itemindent\listparindent
       \rightmargin\leftmargin}\item[\hskip\labelsep
                                     \bfseries Keywords:]}
     {\endlist}

\begin{document}
\maketitle
\thispagestyle{empty}
\pagestyle{plain}

\newcommand{\mfont}{\fontsize{7.5pt}{7.5pt}\selectfont}
\newcommand{\nfont}{\normalsize}

\begin{abstract}
Nested Petri nets have been applied for modeling interaction protocols, mobility, adaptive systems and interorganizational workflows. However, few results have been reported on the use of automated tools for analyzing  the behavior of these nets. In this paper we present a general translation from nested Petri nets into PROMELA and explain how some properties of these nets can be studied using SPIN model checker. Besides, we discuss how to deal with the main limitations that may influence SPIN performance when verifying practical examples.
\end{abstract}


\section{Introduction}
\label{sec:int}

Petri nets (PNs) are one of the most widely used formalisms for analyzing concurrent and distributed systems. The key to their success is the combination of few and simple primitives, a convenient graphical representation and several tools for simulation and verification. Therefore, they have been extended in several ways in order to increase the modeling power. One of the ideas applied to complex models was the use of nesting and recursion. For instance, recursive nets and nets within nets have been used to specify interaction protocols and mobility in the context of distributed multi-agent systems~\cite{Seghrouchni96recursive,Mazouzi:2002,mobil2003,Kissoum:2008}. The nested Petri nets (NPNs) form a representative class of the nets combining these two features. In a NPN, the tokens may be PNs which fire their transitions autonomously or in synchrony with other net tokens~\cite{Lomazova01:fi}. This provides a high degree of modularity and flexibility for the dynamic creation, transportation and removal of concurrent processes. Therefore, its application has been extended to areas such as the coordination of inter-organizational workflows~\cite{iowf-npn} and adaptive systems~\cite{Lomazova08}. NPNs are more powerful than classical PNs and some properties (e.g. reachability and boundedness) are undecidable~\cite{Lomazova00}. However, for important subclasses, such as the multi-level nets and the recursive nested nets with autonomous elements, termination and the inevitability problem can be decided~\cite{Lomazova01:fi}. In spite of this fact, to best of our knowledge, there is no automated tool for analyzing the behavior of NPNs.

In this paper we discuss the use of SPIN for this purpose. SPIN~\cite{spin} is one of most successful tools for simulation and verification of concurrent and distributed software systems. Given a finite-state model of the system behavior, SPIN verifies it against temporal properties by an exhaustive inspection of all possible system states. If some property is violated, a counterexample is provided. SPIN uses a C-like language to specify models called PROMELA (Process Meta Language). Unlike other model checkers, SPIN allows recursive processes; besides, its buffered channels are suitable for implementing the synchronizations in a NPN. The model checking approach suffers from the state-space explosion problem, but clever algorithms have been developed in SPIN to deal with it.

Several variations of PNs and PN-like formalisms (e.g. workflows, business processes, UML diagrams) have been translated into DVE~\cite{Leyla:2010}, LTSA~\cite{Regis2012}, NuSMV~\cite{Eshuis:2006} and SPIN~\cite{Gannod01,Farwer:2004,Ribeiro07,acwfnet2spin,wfnet2spin,Chang:2011}. DVE, LTSA and NuSMV cannot be used in this context because they have no support for recursion. In~\cite{Farwer:2004}, two-level object nets are encoded into Prolog and verified using the XTL model checker. Although the method is intended for arbitrary nesting, the encoding for the synchronization in the multi-level case is not provided. Rewriting logic has been used to express the semantics of recursive algebraic nets~\cite{ECATNets:2008}. But these nets do not include horizontal steps. Translating NPNs into PROMELA is simpler and more amenable for simulation that using rewriting rules or logic programming. Regarding verification, SPIN outperforms Maude model checker in execution time and memory requirements~\cite{Eker:2002}. According to~\cite{Frappier:2010}, SPIN is faster than XTL model checker and can handle a larger number of properties and instances. Among the PROMELA translations, the ones presented in~\cite{Chang:2011,venero2013a} are related to the NPN framework but deal with rather restricted subclasses. The former focus on two-level nested nets without horizontal synchronization or net tokens removal. The later allows to analyze multi-level and recursive NPNs with a rather restricted synchronization and without transportation steps. This paper improves and completes this translation and discusses the effectiveness of SPIN for verifying these nests.

The  remaining sections are organized as follows. The formal definition of NPN that we used in this paper paper is introduced in Section~\ref{sec:npn}. Its translation into PROMELA is presented in Section~\ref{sec:rnpn2promela}. Section~\ref{sec:prop} explains how termination, boundedness and some reachability properties can be studied using SPIN. Section~\ref{sec:performance} analyzes the results obtained on preliminary experiments and discuss the major factors that may influence SPIN performance in practical examples. Section~\ref{sec:conclusion} draws some concluding remarks and future work.

\section{Nested Petri Nets}
\label{sec:npn}

A Petri net~\cite{Murata89} is a 4-tuple $N=(P,T,A,W)$ where $P$ and $T$ are non-empty, finite and disjoint sets of places and transitions resp; $A\subseteq (P\times T)\cup (T\times P)$ is a set of arcs and $W$ is a function defined from $A$ to multisets of uncolored tokens (black dots). A \emph{marking} of a $N$ is a function attaching a multiset of tokens to each place. Transitions represent events (called \emph{firings}) which may change the marking of the net according to $W$. The tokens in a PN have no structure or information. In colored Petri nets (CPNs)~\cite{Jensen92}, each place has a type; thus, it may host tokens with different data values, i.e. colors. The arcs are labeled by multiset expressions containing variables.

A nested Petri net is a CPN in which tokens can also be Petri nets and thus they may fire their own internal transitions~\cite{Lomazova01}. More precisely, a NPN is a tuple $(SN, EN_1,\ldots, EN_n)$ of CPNs, one of them called \emph{system net} ($SN$) and the rest \emph{element nets}. As usual in CPNs, we have a set of basic types $\Sigma$ and a set of basic constants $\Sigma_c$ belonging to these types. In addition, each $EN_i$ is also considered as a constant and a type whose set of values consists of marked net tokens.

The firing of a transition $t$ is performed according to the classical CPN rules. Hence, net tokens can be created, removed and transported as basic ones. In addition,  the firing may synchronized with the firing of other net tokens at the same place (\emph{horizontal synchronization step}) or with the parent net (\emph{vertical synchronization step}). The synchronization is achieved by means of labels that are attached to transitions. The number of tokens that are involved in a horizontal step depends on the label. We have avoided the use of arities for places for simplicity.

In this paper the element nets may share some places of $SN$. This makes the firing of a transition in a net token dependent on the marking of the system net. We restrict the type of these places to net elements with all places of basic types. Besides, in order to avoid conflicts in a synchronizing step, we require that some labeled transitions have no shared place as input.

We assume arc expressions are multisets over (basic or net) constants and variables of fixed type. We denote as ${\it Var}(m)$ the set of variables occurring in the multiset $m$. We also use an anonymous variable (denoted as \_) to indicate a single token, regardless its value. Each occurrence of the symbol \_ represents a different anonymous variable. We use the notation ${\it Var}^*(m)$ for the set of variables and anonymous variables occurring in $m$. The arc inscriptions are further restricted e.g. to avoid testing equality or copying a net token. In the next, we provide  the formal definition and behavior of the NPNs this paper deals with.

\begin{definition}
\label{def:npn}
Let $N= (\Sigma, P_s, L, ar, (EN_0, EN_1,\ldots,EN_b,\ldots, EN_n))$ be a NPN s.t. $\Sigma$ is a finite set of basic types, $P_s$ is a finite set of shared places, $L = L_h\uplus L_v$ is  a sets of labels and $ar: L_h\to \mathbb{N}^{1}$, where $\uplus$ denotes the disjoint union and $\mathbb{N}^{1}$ the natural numbers greater than 1. The set $L_v = L_v^l\uplus L_v^u$ is s.t. $|L_v^l|=|L_v^u|$,

\begin{itemize}
    \item for each $l\in L_v^l$, there is a complementary label $\bar{l}\in L_v^u$ and
    \item for all $l_1,l_2\in L_v^l$, $l_1\neq l_2$ implies $\bar{l}_1\neq \bar{l}_2$.
\end{itemize}

\noindent For all $i=0\ldots n$, $EN_i=(P_i, V_i, C_i, I_i, T_i, \Lambda_i, A_i, W_i)$ is a colored Petri net, called net component: $EN_0$ is called the \emph{system net}, denoted as $SN$, and the remaining are called\emph{ element nets}. Each each $EN_i$ we have that

\begin{enumerate}
    \item $P_i$ is a finite set of places s.t. $P_s\subset P_i$ if $i=0$ and $P_i\cap P_s=\emptyset$ if $i>0$,
    \item $V_i$ is a set of variables,
    \item $C_i$ is a type function s.t. if $0<i\leq b$ then  $C_i:P_i\cup V_i \to \Sigma$ otherwise $C_i:P_i\cup V_i \to \Sigma\cup \mathcal{P}(\{EN_1,\ldots, EN_n\})$. Besides, for all $p\in P_s$, $C_0(p)\in\Sigma\cup\mathcal{P}(\{EN_1,\ldots, EN_b\})$.
    \item $I_i$ is function defined from $P_i$ into multisets over $\Sigma_c\cup\{EN_1,\ldots, EN_b\}$,
    \item $T_i$ is a finite set of transitions s.t. $P_i\cap T_i=\emptyset$,
    \item $\Lambda_i:T_i\to L\cup\{\epsilon\}$ is the labeling function. The symbol $\epsilon$ denotes the empty label and it is used for unlabeled transitions. For $SN$ we have that $\Lambda_0:T_0\to L_v^l\cup\{\epsilon\}$,
    \item $A_i$ is a set of arcs s.t. if $0<i\leq b$ then $A_i\subseteq (P_i\times T_i) \cup (T_i \times P_i)$ otherwise $A_i\subseteq ((P_s\cup P_i) \times T_i) \cup (T_i \times (P_s\cup P_i))$. Besides, for all $(p,t)\in A_i$, if $\Lambda_i(t)\in L_h\cup L_v^u$ then $p\in P_i$,
    \item $W_i$ is an arc expression function defined from $A_i$ to multisets over $V_i\cup\Sigma_c\cup\{EN_1,\ldots, EN_n\}$ s.t.
    \begin{enumerate}
        \item[(a)] there are no net constants in input arc expressions;
        \item[(b)] every variable has at most one occurrence in each input arc expression. In addition, every net variable has at most one occurrence in each output arc expression;
        \item[(c)] given two different input arcs of the same transition $(p_1, t)$ and $(p_2, t)$, ${\it Var}(W_i(p_1, t))\cap {\it Var}(W_i(p_2, t))=\emptyset$;
        \item[(d)] for each variable $x\in {\it Var}(W_i(t, q))$, there should be an input arc of $t$ s.t. $x\in {\it Var}(W_i(p, t))$.
        If $C_i(x)\notin \Sigma$ then there is no different output arc $(t, q')$ s.t.  $x\in {\it Var}(W_i(t,q'))$;
        \item[(e)] there are no net-typed anonymous variables in output arc expressions.
    \end{enumerate}
\end{enumerate}

\end{definition}

The places in a NPN can be divided into two kinds, places with basic type or net type. All places in the net components $EN_1,\ldots,EN_b$ are of basic types. Due to condition on $C_0$ in the above definition, in this paper, the shared places can be of basic or 1-level types (although multi-level nets may also be allowed). We assume the type of each constant element in the multiset of an input (output) arc must be included in the type of the corresponding input (output) place. The type of a variable occurring in an arc inscription must coincide with the type of the incident place.

\begin{example}
\label{ex:npn-fact}
As running example we use a NPN, adapted from~\cite{Lomazova01}, that it is shown in Figure~\ref{figRPNFact}. Places are drawn as ellipses and transitions as bars. We omit the arc labels $\{1\}$ as well as the braces for multisets of a single element. The NPN has two net components $SN$ and the element net $F$ which simulates the recursive calls for computing the factorial function. Here $F$ shares two places from $SN$: $p1$ which initially may store $a\geq 0$ black tokens and $p5$ which is initially empty. The places $p3$ and $p7$ are net-typed while the rest are uncolored. The net $SN$ simulates the computation of the factorial of an integer $0\leq b\leq a$.

\vspace{-10pt}
\begin{figure}
\begin{center}
\includegraphics[scale=0.85]{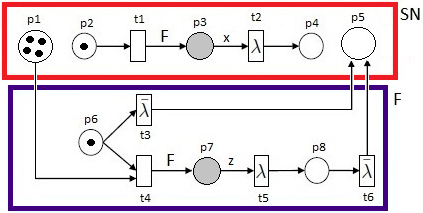}
\end{center}
\vspace{-10pt}
\caption{Example of NPN for simulating the factorial computation calls.}
\label{figRPNFact}
\end{figure}
\end{example}

\subsection{State of a NPN}

The state or configuration of PN, called \emph{marking}, is a distribution of tokens over the places. In a NPN, the net tokens have their own markings. Hence, a \emph{marking of an element net} $EN_i$ over $N$, $1\leq i \leq n$, is inductively defined as follows.
\begin{enumerate}
    \item A function $M$, mapping each place $p\in P_i$ to a finite multiset over $\Sigma$ is a marking of $EN_i$ over $N$. The pair $(EN_i, M)$ is called a marked element net or net token of $EN_i$.
    \item Let $\bar{\Sigma}$ be a set of marked element nets. Then, a function mapping each place $p\in P_i$ to a finite multiset over $\bar{\Sigma}\cup \Sigma$, is also marking of $EN_i$ over $N$.
\end{enumerate}

The marking of $EN_0$  is a function mapping each place $p\in P_0$ to a finite multiset over $\bar{\Sigma}\cup\Sigma$. A \emph{marking of a NPN} $N$ is a marking of $SN$.

Any marking must match the type definition of the places. Hence, for all $p\in P_i$, if $C_i(p)\in\Sigma$, then $M(p)$ is a multiset over $C_i(p)$; otherwise $M(p)$ is a multiset of net tokens of $C_i(p)$.  To avoid confusion, it can be assumed that places, transitions, variables and arcs of two net tokens of the same element net are different. Notice that, no net token of $SN$ is allowed and places belonging to $P_s$ are shared by all net tokens of the net components $EN_{b+1},\ldots, EN_n$.

\begin{figure}
\begin{center}
\includegraphics[scale=0.85]{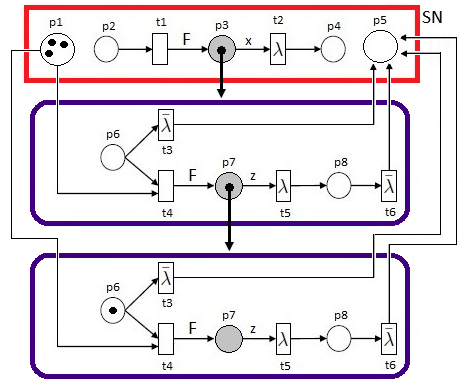}
\end{center}
\vspace{-10pt}
\caption{A marking for a NPN.}
\label{figRPNFactMarking}
\end{figure}

The \emph{initial marking} of any net component is obtained from the initial function $I_i$. By definition, this function has no net token of type $EN_{b+1},\ldots, EN_n$. The initial marking of $N$, obtained from $I_0$, is denoted as $M_0$. For all $i>0$, $EN_i$ also represents a constant corresponding to the marked net $(EN_i,I_i)$. Figure~\ref{figRPNFactMarking} shows a marking for the NPN in Example~\ref{ex:npn-fact}. The two nested net tokens share the places $p_1$ and $p_5$ of $SN$.

In the later we will say that a net token $nt'=(EN_i,M')$ occurs in marking $M$ if there is a place $p$ s.t. either $nt'\in M(p)$  or there exists $(EN_j,M'')\in M(p)$ and $nt'$ occurs in $M''$. The occurrence of two net tokens at the same place in a marking is defined analogously. The replacement of $nt'$ in $M$ by a net token $nt_1'=(EN_i,M_1')$ (denoted as $M[nt'\to nt_1']$) is defined as the marking $M_1$ s.t. $M_1(p)=M(p)-\{nt'\}\cup \{nt'_1\}$ if $nt'\in M(p)$; $M_1(p) = M(p)-\{(EN_j,M'')\}\cup \{(EN_j,M''[nt'\to nt_1'])\}$ if $nt'$ occurs in $M''$; and $M_1(p') = M(p')$ otherwise.

\subsection{Behavior of a NPN}

The behavior of a PN is defined by the sequences of steps that it may execute from an initial state. In a NPN each step may involve the firing of several transitions. As usual in CPNs, each firing is conditioned to the binding of the variables in the input arcs. These concepts are defined below.

\noindent\textbf{Binding\ }  Let $t$ be a transition in a net component $EN=(P, C, I, T, \Lambda, A, W, V)$ of a NPN. We write ${\it Var}^*(t)$ for the set of all variables occurring in input arcs of $t$. Hereafter, we assume that $W(p,t)$ (resp. $W(t,p)$) is the empty set if $(p,t)\notin A$ (resp. $(t,p)\notin A$). A \emph{binding} for $t$ is a function $b$ assigning to each variable $x \in {\it Var}^*(t)$ a value from $\bar{\Sigma}\cup\Sigma$ (of the corresponding type). It is applied to multisets in a straightforward way. The set $\{b(x)\mid x\in {\it Var}^*(t) \wedge C(x)\notin\Sigma\}$ are the net tokens involved in $t$ w.r.t. $b$.

\noindent\textbf{Firing\ }  Let $M$ be marking of a NPN $N$. A transition $t\in T_0$ is \emph{enabled} in $M$ w.r.t. a binding $b$, if for all $a=(p, t)\in A_0$, $b(W_0(a))\subseteq M(p)$. In this case, $t$  may fire. After the firing, a new marking $M_n$ is obtained s.t. for any place $p\in P_0$, $M_n(p)=M(p)-b(W_0(p,t))\cup b(W_0(t,p))$.

Let $(EN_i,M')$ be a net token occurring in $M$ at some place $p'$. A transition $t$ of $(EN_i,M')$ is \emph{enabled} in $M$ w.r.t. a binding $b$, if for all $a=(p, t)\in A_i$, $b(W_i(a))\subseteq M'(p)$. If $t$ fires, a new marking $M''$ is obtained from $M'$ s.t. for any place $p\in P_i$, $M''(p)=M'(p)-b(W_i(p,t))\cup b(W_i(t,p))$. Furthermore, a new marking $M_n$ is obtained from $M$ s.t. for any place $p\in P_s$, $M_n(p)=M(p)\cup b(W_i(t,p))$; for any place $p\notin P_s\cup\{p'\}$, $M_n(p)=M(p)$; and $M_n(p')=M[(EN_i,M')\to (EN_i,M'')](p')$.

\noindent\textbf{Step\ } A NPN allows autonomous and synchronizing steps; the latter divided into horizontal and vertical. An \emph{autonomous step} is the firing of a single unlabeled transition in $SN$ or in a net token of $N$. This step is denoted as $M[t\rangle M_n$ where $M_n$ is the resulting marking.  A \emph{horizontal} step is the firing of $k$ transitions labeled as $l\in L_h$ with $ar(l)=k$, in $k$ different net tokens that occur in $M$ at the same place. This step is denoted as $M[t_1\ldots t_k\rangle M_n$. A \emph{vertical step} is the firing of a transition $t$ in ${\it SN}$ or some net token that occurs in $M$ s.t. $l=\Lambda(t)\in L_v^l$, and the firing of a transition labeled as $\bar{l}$ in all net tokens involved in the binding of $t$. This step is also denoted as $M[t\rangle M_n$ but the notation may also include the transitions fired in the child nets.

We remark that transitions involved in a synchronizing step do not share input places, thus the order for firing them is irrelevant. The vertical synchronization is performed between two adjacent levels of nesting: transitions in $L_v^l$ are intended for \emph{lower} synchronization (with child net tokens) while those in $L_v^u$ are for \emph{upper} synchronization (with the parent). 

A marking $M$ is called \emph{reachable} if there is a sequence of zero or more steps  $M_0[\rangle M_1 [\rangle \ldots[\rangle M_k$ s.t. $M_k=M$. This is denoted as $M_0[*\rangle M$. It is called \emph{dead} if no step can be done from it.  A NPN terminates if it has no infinite firing sequence.  The net has a cycle if there is a reachable marking $M$ s.t. $M_0[*\rangle M[*\rangle M$.

\begin{example}
The firing sequence below corresponds to the NPN in Figure~\ref{figRPNFact}. We write a marking as a sequence of pairs $p:M(p)$ enclosed by the symbols $\lessdot$ and $\gtrdot$. Uncolored places are marked with non-negative integers instead of multisets of black dots. We use superscripts for the places and transitions in net tokens.

\vspace{7pt}
$\lessdot p_1:4,p_2:1,p_3:\emptyset, p_4:0,p_5:0\gtrdot\ \mathbf{[t_1\rangle}$

$\lessdot p_1:4,p_2:0, p_3:(F^1,\lessdot p_6^1:1,p_7^1:\emptyset,p_8^1:0\gtrdot),p_4:0,p_5:0\gtrdot\ \mathbf{[t^1_4\rangle}\ $

$\lessdot p_1:3,p_2:0, p_3:(F^1,\lessdot p_6^1:0,p_7^1:(F^2,\lessdot p_6^2:1,p_7^2:\emptyset,p_8^2:0\gtrdot),p_8^1:0\gtrdot),$

\verb"           " $\ p_4:0,p_5:0\gtrdot\ \mathbf{[t^2_3\ t^1_5\rangle}\ $

$\lessdot p_1:3, p_2:0, p_3:(F^1,\lessdot p_6^1:0,p_7^1:\emptyset,p_8^1:1\gtrdot),p_4:0,p_5:1\gtrdot\ \mathbf{[t^1_6\ t_2\rangle}$

$\lessdot p_1:3,p_2:0,p_3:\emptyset, p_4:1,p_5:2\gtrdot $
\vspace{7pt}

The two first steps of this sequence create two nested net tokens (say $nt_1$ and $nt_2$) at $p_3$ and $p_7^1$ resp. After that, the inner net token performs a vertical step with its parent, firing the transitions $t^2_3$ and $t^1_5$ resp. This step adds a black dot at $p_5$ and $p_8^1$ and consumes $nt_2$. Another vertical step occurs between $nt_1$ and $SN$ involving the transitions $t^1_6$ and $t_2$ resp. As before a black dot is added at $p_5$ and also at $p_4$; besides $nt_1$ is consumed. Then, no further step can be done in the net. An alternative sequence is obtained if in the second step above, we choose the transition $t_3^1$ instead of $t_4^1$. In this case, the sequence is the next:

\vspace{7pt}
$\lessdot p_1:4,p_2:1,p_3:\emptyset, p_4:0,p_5:0\gtrdot\ \mathbf{[t_1\rangle}$

$\lessdot p_1:4,p_2:0, p_3:(F,\lessdot p_6^1:1,p_7^1:\emptyset,p_8^1:0\gtrdot),p_4:0,p_5:0\gtrdot\ \mathbf{[t^1_3\ t_2\rangle}$

$\lessdot p_1:4,p_2:0,p_3:\emptyset, p_4:1,p_5:1\gtrdot $
\vspace{7pt}

In general, when the net reaches a dead marking $M$ we have $M(p_4)=1$, $M(p_5)=b+1$ for some $0\leq b\leq a$ (i.e.  the number of net tokens - factorial calls), $M(p_1)=a-b$, $M(p_2)=0$ and  $M(p_3)=\emptyset$.
\end{example}

\section{Translating Nested Petri Nets into PROMELA}
\label{sec:rnpn2promela}

In this section we explain how to translate a NPN into PROMELA. PROMELA basic elements are variables, processes and  (synchronous or asynchronous) channels. Processes are specified using imperative statements such as assignments, conditional constructions, loops and communications actions. Process communication may be performed using messages which can be read from and written to channels. Several instances of processes may be running concurrently when executing a PROMELA program. The execution of a process may be interrupted by another process, unless the sequence is defined as an atomic or a deterministic region. Hereafter, we  assume the reader is familiar with the basics of PROMELA  and  SPIN  semantics of executability (see e.g.~\cite{spin}).

In our translation we represent each element net using a \small\verb"proctype"\nfont\ definition; thus each net token is a process.  We assume that each transition has an identity number represented in a byte and that $L\subseteq [1, 254]$. We denote the label of a transition $t$ as \verb"L(t)" In addition, we use \verb"Lvl(t)" to indicate the label for lower synchronization and \verb"-Lvl(t)" for  its complementary. For transitions with labels in $L_v^u\cup L_h\cup\{\epsilon\}$, we define \verb"Lvl(t)=0" and \verb"-Lvl(t)=255". 

We assume that basic constants are represented as integer values. Each colored basic place is translated into an asynchronous channel which stores the tokens. For simplicity, we use a basic representation of this channel; hence, it may contain multiple occurrences of the same token\footnote{An improved version may use an additional field for the multiplicity of the token.}. Uncolored places are best represented as non-negative integer variables.  As an alternative, a colored basic place may be unfolded into several uncolored ones.

A net place is also represented as channel that in addition  is used for exchanging messages with the processes corresponding to net tokens at the place. Each message consists of five fields. The first holds the instantiation number (\small\verb"_pid"\nfont) of the net token process sending or receiving the message. The second and the third fields are the label and the identity number of the transition which is enabled, resp. The four bit field defines the type of the message: a synchronization request (0) or a response (1). In a response message, the last field indicates whether or not the net token is consumed after the step. Additional fields may be used depending on the application, e.g. for encoding the net type or exchanging data between net tokens.

The general structure of the translation is shown in Figure~\ref{figStructSN}. The system net is translated as the \verb"init" process. Shared places must be declared as global variables while non-shared ones are local to the process definition of the net component. Arc variables of basic type are also translated as local variables while net-typed variables are omitted. The initial marking for uncolored places can be defined as the initialization part of the variable declaration. For colored places, we may use an atomic region for producing each token at the corresponding place as explained below.

\mfont
\begin{figure}
\begin{center}
\begin{verbatim}
 1   typedef BasicPlace { chan d = [MaxTok] of {byte} }
 2   typedef NetPlace { chan d = [MaxMsg] of {byte,byte,byte,bit,bit} }
 3   chan gbChan = [MaxMsg] of {byte, chan, byte, byte, bit};
 4
 5    /* Auxiliary Code, Shared Places, Element Nets */
 6
 7   init(){
 8   /* Non-Shared Places and Arc Variables */
 9   atomic{ set_priority(0,2); /* Initial Marking */; set_priority(0,1) }
10   do:: d_step{ enableTest_t ->
11            set_priority(0, 6);
12            consumeActions_t; produceActions_t;
13            set_priority(0, 1) }
14     :: ...
15   od }
\end{verbatim}
\end{center}
\vspace{-5pt}
\caption{General structure of the PROMELA specification for a NPN.}
\label{figStructSN}
\end{figure}
\nfont

The behavior of PN is usually translated into PROMELA using \small\verb"do"\nfont-loop where each option simulates the firing of a transition~\cite{spin}. The options have the form\\  \small\verb"atomic{enableTest -> consumeActions; produceActions}"\nfont. Nevertheless, we have introduced some variations to this schema to effectively simulate the multiple firings in a NPN step and also to improve efficiency during verification. First, we have chosen the use of deterministic sequences instead of atomic regions. Since a process instance cannot be created inside a \small\verb"d_step"\nfont, the atomic sequence must be used for transitions that create net tokens. In addition, we use priorities to avoid the interleaving of firings that do not belong to the step.

\subsection{General rules for translating the transitions}

For each transition $t$ in $SN$ or an element net, the expression \small\verb"enableTest_t"\nfont\  is the conjunction of an enabling condition for each input arc $(p,t)$. This condition, denoted as \small\verb"enableTest"\nfont$_p$, depends on the arc inscription $W(p,t)$. Analogously, \small\verb"consumeActions_t"\nfont\ is the sequence of instructions for removing the tokens on each input label $W(p,t)$. Finally, \small\verb"produceActions_t"\nfont\ is the sequence of instructions for adding the tokens in $W(t,p)$ to each output place $p$.

W.l.o.g. we assume that each incident arc to an uncolored place $up$ is labeled by a natural number. Hence, if $W(up,t) = n$ then \small\verb"enableTest"\nfont$_{up}$ is the expression \small\verb"up >= n"\nfont\ and \small\verb"consumeActions"\nfont$_{up}$ is the instruction \small\verb"up = up - n"\nfont. If $W(t, up) = n$, \small\verb"produceActions"\nfont$_{up}$ is defined as \small\verb"up = up + n"\nfont. An incident arc to a colored place is labeled by a multiset. Therefore, for a place $bp$ of basic type, \small\verb"enableTest"\nfont$_{bp}$ may also be a conjunction having the form \small\verb"len(bp.d) >= n && E(c1) ... && E(cm)"\nfont, where  $n=|W(bp,t)|$ and  \small\verb"E(ci)"\nfont\  is an enabling expression for each basic constant in $W(bp,t)$. The expression \small\verb"E(ci)"\nfont\  depends on the multiplicity $k=|W(bp,t)|_{ci}$ of the constant $ci$ and it is defined as

$$ \small\verb!E(ci)!\normalsize = \left\{
  \begin{array}{l l}
    \small\verb!bp.d ?? [ci]!\normalsize & \quad \text{if $k=1$}\\
    \small\verb!c_expr{ numTok(qptr(PProcName->bp.d), ci)>=k }!\normalsize  & \quad \text{if $k>1$}
  \end{array} \right.$$

\noindent For a net place $np$, the multiset $W(np, t)$ is composed of distinct variables. Hence,  \small\verb"enableTest"\nfont$_{np}$ must check the existence of $k=|W(np,t)|$ request messages at the channel, with the complementary label of $t$. It is defined as follows

$$\left\{
  \begin{array}{l l}
    \small\verb!np.d ?? [_,-Lvl(t),_,0,0]!\normalsize & \quad \text{if $k=1$}\\
    \small\verb!c_expr{ numMsg(qptr(PProcName->np.d-1), -Lvl(t))>=k }!\normalsize  & \quad \text{if $k>1$}
  \end{array} \right.$$

\noindent In the above expressions we have used SPIN facility for embedding C code into PROMELA models. The C functions  \small\verb"numTok"\nfont\ and \small\verb"numMsg"\nfont\  compute the number of tokens and messages in a channel corresponding to a basic or net place respectively (see  Appendix~\ref{app:aux.code} for the definition and further details).

For a colored place $p$, \small\verb"consumeActions"\nfont$_p$ is a sequence of instructions for removing from $p$ the tokens in $b(W(p,t))$. Similarly, \small\verb"produceActions"\nfont$_p$ is the sequence of instructions for adding the tokens in $b(W(t,p))$ to $p$. Table~\ref{tab.act.bas.places} shows these PROMELA instructions when $p$ has basic type.
W.l.o.g. we assume that all constants appear fist in $W(p,t)$, because it is easier for computing the binding. We also assume that the name of an arc variable and the corresponding PROMELA variable are the same. Since the channels represent multisets, the tokens or messages should be non-deterministically inserted or removed. In this paper we chose the latter option. The macro definition for removing a message from a net channel is provided in Appendix~\ref{app:aux.code}; the one for a basic channel is analogous. For the sake of readability, we use a notation similar to the standard SPIN statement for receiving messages, i.e. \small\verb"chan ?* msg"\nfont. In the last cell of the table, we assume \small\verb"v1 ...vn"\nfont\ are all the values belonging to the type of the output place.

\small

\begin{table}[h!]
\begin{center}\begin{tabular}{|c|l|l|l|}
      \hline
      \verb" Arc label "  & \verb" consumeAction " & \verb" produceAction " \\
      \hline
      \verb"c"  & \verb" bp.d ?? c "  & \verb" bp.d ! c " \\
      \hline
      \verb"x"  & \verb" bp.d ?* x "  & \verb" bp.d ! x "\\
      \hline
      \verb"_"  & \verb" bp.d ?* _ "  & \verb" if :: bp.d!v1  ...  :: bp.d!vn  fi  "\\
      \hline
\end{tabular}\end{center}
\vspace{-2pt}
\caption{Consume and produce actions for basic-typed arc labels.}
\label{tab.act.bas.places}
\end{table}
\normalsize

Table~\ref{tab.act.net.places} shows the PROMELA instructions for dealing with net-typed labels. For a net place $np$, the label of an output arc $W(t,np)$ includes just net constants; hence, the instructions in \small\verb"produceActions"\nfont$_{np}$ create a net token for each constant in $W(t,np)$. As shown in the last row, a child net token $nt$ is produced at a place $np$ by creating a process instance for the corresponding element net. The process instantiation number identifies the net token messages. The initial message sent to the the channel, represents the net token at the place. Besides, it allows the removal of $nt$ without  synchronization, for instance by an unlabeled transition or when the parent net is consumed. The process is created with priority 2 in order to compute the initial marking of the $nt$.

\small
\begin{table}[t!]
\begin{center}\begin{tabular}{|l|l|}
      \hline
      \verb" consumeAction(_)" & \verb" np.d ?* nt,-Lvl(t),it,0,0; "    \\
                               & \verb" consNetTok(np.d,nt);"          \\
                               & \verb" np.d ! nt,-Lvl(t),it,1,1;"        \\
                               & \verb" if :: Lvl(t)==0 -> set_priority(nt, 3)" \\
                               & \verb"    :: else      -> set_priority(nt, 5)"  \\
                               & \verb" fi"                   \\
      \hline
      \verb" transportAction(x) "        & \verb" np.d ?* nt,-Lvl(t),it,0,0;"\\
      \verb"   np == opn && Lvl(t)>0  "  & \verb" np.d ! nt,-Lvl(t),it,1,0;"   \\
                                         & \verb" set_priority(nt, 5)"        \\
      \hline
      \verb" transportAction(x)" & \verb" np.d ?* nt,-Lvl(t),it,0,0;"\\
      \verb"   np != opn "         & \verb" transpNetTok(np.d,onp.d,nt);" \\
                                 & \verb" gbChan ! nt,opn.d,-Lvl(t),it;"    \\
                                 & \verb" if :: Lvl(t)==0 -> set_priority(nt, 3);  " \\
                                 & \verb"                    onp.d ! nt,255,0,0,0;"  \\
                                 & \verb"    :: else      -> set_priority(nt, 5);"   \\
                                 & \verb" fi"                   \\
      \hline
      \verb" produceAction(EN_i)" & \verb" nt = run ElementNet_i(np.d) priority 2;"\\
                                  & \verb" np.d ! nt,255,0,0,0 "\\
      \hline
\end{tabular}\end{center}
\vspace{-2pt}
\caption{Consume, produce and transport actions for net-typed arc labels.}
\label{tab.act.net.places}
\end{table}
\normalsize

As we explained above, a transition $t$ with an arc $(np, t)$ may consume or transport $nt$ if the channel contains a request message from $nt$, with the complementary label of $t$.  When $t$ fires, \small\verb"consumeActions"\nfont$_{np}$ removes the remaining request messages of $nt$ (\small\verb"consNetTok"\nfont) from the channel. Then, a response message is sent that will force the termination of the process $nt$. If $nt$ is transported to a new place $onp$, then \small\verb"transportActions"\nfont$_{np}$ moves the messages to the new channel (\small\verb"transpNetTok"\nfont). In this case, the response message is sent to $nt$ by means of a global channel (\small\verb"gbChan"\nfont). If the places $np$ and $onp$ coincide and the $t$ is labeled for vertical synchronization\footnote{No translation is required in case $np=onp$ and $t$ is unlabeled.}, then the response message is sent via \small\verb"ppChan"\nfont\  with the indication that the $nt$ should not be removed.  In all cases, by changing the priority of $nt$ we ensure that once $t$ completes the firing, the child net processes will be executed in order to conclude the synchronization. The priority is lower for those net tokens that are consumed or transported to a new place without synchronization. Finally, we point out that other features of CPNs such as guards for transitions or inhibitor and reset arcs, can be easily embedded into the translation; as well as additional actions such as printing the marking and binding.

\subsection{Translating the element nets}

Each element net is translated into a \small\verb"proctype"\nfont\ definition having as input parameter the channel corresponding to the place at the parent net where the net token was created. Its behavior is also simulated using two nested \small\verb"do"\nfont-loops, as shown in Figure~\ref{figStructNetEl}. The inner loop includes an option for each transition $t$ that may execute either an autonomous firing (if $\Lambda_i(t) \in L_v^l\cup \{\epsilon\}$) or a request for synchronization (if $\Lambda_i(t) \in L_v^u\cup L_h$). In addition, it also includes an option for each label in $L_h$, in order to set up a horizontal synchronization. This cycle is broken  once a response message is received, by means of an \small\verb"unless"\nfont\  instruction. Since the \small\verb"unless"\nfont\  interrupts the atomic sequences, all options in the inner loop must be enclosed in a \small\verb"d_step"\nfont.

\mfont
\begin{figure}[t!]
\begin{verbatim}
 1  proctype EN_i(chan ppChan){
 2  /* Non-Shared Places and Arc Variables */
 3  atomic{ /* Initial Marking */; set_priority(_pid, 1) }
 4  do:: { do:: d_step{/* autonomous firing          - Op1 or Op2 */}
 5           :: d_step{/* synchronization request    - Op3* /}
 6           :: d_step{/* horizontal synchronization - Op4 */}
 7           :: ...
 8          od }
 9        unless atomic{ gbChan ?? [eval(_pid),ppChan,lt,it]
10                   ||  ppChan ?? [eval(_pid),lt,it,1,rm]       ->
11          if:: gbChan ?? [eval(_pid),pc,lt,it] ->
12               gbChan ?? eval(_pid),pc,lt,it; rm = 0;
13            :: else -> pc ?? eval(_pid),lt,it,1,rm;
14          fi;
15          if:: it==idt(t) ->                           /* aut step */
16               produceActions_t; rmConf_t;
17            :: lt==L(t) && enableTest_t ->             /* sync step */
18               consumeActions_t; produceActions_t; rmConf_t;
19            :: lt==255 -> skip
20          fi;
21          if:: rm -> break ::else -> set_priority(_pid, 1) fi }
22  od;
23  d_step{ consNetsAtPlace(np1); ... }
24  set_priority(_pid, 1) }
\end{verbatim}
\vspace{-5pt}
\caption{PROMELA specification for an element net.}
\label{figStructNetEl}
\end{figure}
\nfont

A transition $t$ with label in  $L_v^l\cup \{\epsilon\}$ is translated similarly to Figure~\ref{figStructSN}. However, there are two situations where the translation may differ. Firstly, if $t$ is in conflict\footnote{ A transition $t$ is in conflict with another transition $t'$ if there is a reachable marking $M$ s.t. both transitions are enabled and the firing of $t$ disables $t'$.} with another transition with label in $L_h\cup L_v^u$, its translation should include an additional code \small\verb"rmConf_t"\nfont\ that we explain at the end of this section. Besides, if $t$ creates net tokens, the new processes cannot be created inside the \small\verb"d_step"\nfont. Therefore, if $t$ does not produce net tokens, it is translated as \small\verb"Op1"\nfont\  in Table~\ref{tab.opt.el.net}. Otherwise, the firing is split into two parts. The first part, i.e. the removal of the input tokens, takes place inside the inner loop and it is translated as \small\verb"Op2"\nfont\   in Table~\ref{tab.opt.el.net}. Note that, the last instruction sends a response message  with the identity of the transition to terminate the inner cycle. Hence, the other part of the firing (for producing the output tokens\footnote{The second part of the firing can be restricted just to creation of net tokens; the other actions may remain in the inner cycle part.}) is executed at the outer cycle as an \small\verb"if"\nfont\ option, as shown in Figure~\ref{figStructNetEl}, lines \small\verb"11-12"\nfont.


\small
\begin{table}[t!]
\begin{center}\begin{tabular}{|l|l|}
      \hline
      \verb" Op1 "  & \verb" enableTest_t ->  "  \\
                    & \verb"       set_priority(_pid, 6);"  \\
                    & \verb"       consumeActions_t;"  \\
                    & \verb"       produceActions_t; rmConf_t;"  \\
                    & \verb"       set_priority(_pid, 1);"  \\
      \hline
      \verb" Op2 "  & \verb" enableTest_t -> set_priority(_pid, 6); "  \\
                    & \verb"       consumeActions_t;" \\
                    & \verb"       ppChan!_pid,0,t,1,0"  \\
      \hline
      \verb" Op3 "  & \verb" enableTest_t && ! ppChan??[eval(_pid),L(t),_,0,0] ->  "\\
                    & \verb"       ppChan!_pid,L(t),t,0,0" \\
      \hline
      \verb" Op4 "  & \verb" ppChan??[eval(_pid),Lh,_,0,0] && "  \\
                    & \verb" c_expr { numMsg(qptr(PEN_i->ppChan-1), Lh) >= ar(Lh) } ->  "  \\
                    & \verb"       set_priority(_pid, 6); "  \\
                    & \verb"       ppChan ?? eval(_pid),Lh,it,0,0; ppChan ! _pid,Lh,it,1,0; "  \\
                    & \verb"       /* repeat the next code ar(Lh)-1 times */ "  \\
                    & \verb"       ppChan ?* nt,Lh,it,0,0; ppChan ! nt,Lh,it,1,0; "  \\
                    & \verb"       set_priority(nt, 4); "  \\
      \hline
\end{tabular}\end{center}
\caption{Different options in the inner loop of an element net.}
\label{tab.opt.el.net}
\end{table}
\normalsize

Transitions labeled for upper or horizontal synchronization depends on the other net tokens to fire. Due to this, its translation is divided into two parts. The first is performed once the transition is enabled and it just sends a request message to the parent. The option in the inner loop is as \small\verb"Op3"\nfont\  in Table~\ref{tab.opt.el.net}. The second part (actually the firing) is performed once the process receives the parent response message, as shown in Figure~\ref{figStructNetEl}, lines \small\verb"17-18"\nfont. Note that, any enabled transition with the same label may be chosen for firing.

A net token may be transported or consumed as result of a vertical synchronization. In the former case, the response message is received via \small\verb"gbChan"\nfont, along with its new location (\small\verb"ppChan"\nfont). If the net was consumed by the parent, the message is received via \small\verb"ppChan"\nfont\  and its last field is 1. Therefore, the outer loop is broken and the child nets that may be still active at some place are removed without synchronization. The last \small\verb"d_step"\nfont\ (line \small\verb"23"\nfont) deals with this situation.
It also applies when the net is consumed by an unlabeled transition.  As Figure~\ref{figStructNetEl}, line \small\verb"19"\nfont\ shows, no action is performed when the net is transported or consumed without synchronization.

All net tokens involved in a horizontal step are situated at the same place and remain there after the firing. Hence, the synchronization may be arranged by any of the possible participants (any net token with an enabled horizontal transition). To this end, the inner loop should include an option for each horizontal label occurring in the element net.  This option checks if the number of requests at \small\verb"ppChan"\nfont\  fulfills the arity of the label. If so the remaining participants are non-deterministically chosen and the response messages are sent via \small\verb"ppChan"\nfont. The PROMELA code is shown in the last row of Table~\ref{tab.opt.el.net}. Just a slight modification is required if the arities are attached to places instead of labels. In this respect, the arity of the place where the net token is created should be an additional parameter of the \small\verb"proctype"\nfont\  definition. As the parent channel, this parameter should be updated whenever the net token is transported to a new place.

We  remark that the rules for generating \small\verb"enableTest"\nfont, \small\verb"consumeActions"\nfont\ and \small\verb"produceActions"\nfont\ are the same for all transitions, regardless the label. Note that when a transition $t$ with label in $L_h\cup L_v^u$ consumes net tokens, the label used for \small\verb"enable"\nfont\  and \small\verb"consume"\nfont\ is the one for lower synchronization, \small\verb"Lvl(t)=0"\nfont\ as for unlabeled transitions. Therefore, the translation of the firing is the same. For these transitions an additional situation should be considered, i.e. when they are disabled by the firing of another transition $t1$\footnote{Transitions $t$ and $t1$ belong to the same net token because $t$ has no shared input place.}. In this case, if there is a synchronizing request from $t$ at the parent channel, then it should be removed. To deal with such a conflict, each transition $t1$ having a common input place (or output place if e.g. inhibitor arcs are allowed) with $t$ must include the next code as part of \small\verb"rmConf_t1"\nfont. For simplicity, this code just removes the request message without checking whether $t$ was disabled. The request can be rewritten later, in case $t$ (or any other transition with the same label) remains enabled after firing $t1$.

\small
\begin{verbatim}
if :: ppChan ?? [eval(_pid),_,t,0,0] -> ppChan ?? eval(_pid),_,t,0,0
   :: else fi
\end{verbatim}
\nfont


\subsection{Correctness issues}

In the later we argue about the correctness of the above translation schema. Our first argument is the fact that places, tokens, labels, variables and input/output arcs are effectively translated, and the firing of a transition is properly computed. Besides, after the initializing the \small\verb"init"\nfont\  process (Figure~\ref{figStructSN}, line \small\verb"9"\nfont), the state of the PROMELA program will correspond to the initial marking of the net.  Hence, if a transition is enabled in the initial marking, it will also be enabled in the PROMELA program.  A step $M_0[\rangle M_1$ in the net will correspond to a sequence of atomic or deterministic regions that are executed with priorities between 6 and 2. In an autonomous or vertical step, the first region can be \small\verb"atomic"\nfont\  or \small\verb"d_step"\nfont\ (depending whether or not the process is \small\verb"init"\nfont) and it is always executed with priority 6. If the firing is split, then an additional \small\verb"d_step"\nfont\   of priority 6 follows the first one. In a vertical step, the sequence continues with several \small\verb"atomic"\nfont\  with priority 5, corresponding to the firing of the upper transitions involved in the synchronization. In a horizontal step, the sequence starts with a \small\verb"d_step"\nfont\  of priority 6, that selects the participant net tokens. The first firing of the horizontal synchronization is also executed in a \small\verb"d_step"\nfont\   with priority 6 while the remaining with priority 4.

In all cases, the previous sequence may be followed a number of regions with priority 3 corresponding to those net tokens that are transported or consumed without synchronization. It may finish with \small\verb"d_step"\nfont-s  with priority 2 corresponding to the initialization of the net tokens that are created in the step. By choosing the corresponding processes and branches, the state of the PROMELA program after completing such sequence will match the marking $M_1$. By the SPIN  semantics, processes having the same priority are interleaved. Hence, several execution sequences may correspond to the same step in the NPN. Since concurrent events are interleaved in all possible ways, we guarantee that all possible steps from an initial marking can be reproduced by the PROMELA translation.

Before the sequence described above, the translation may perform \small\verb"d_step"\nfont-s with priorities 1 corresponding to synchronization requests. However, the requests are executed once before the synchronization, unless they are disabled by another firing. Besides, at the channel there will be a single request message per net token per label. Therefore, the number of such  \small\verb"d_step"\nfont-s is finite. These regions modify the state of the PROMELA program but do not affect the underlying marking\footnote{The same applies to other PROMELA steps for the control flow of the program.}. Hence, any firing sequence in the NPN can be simulated using the PROMELA translation, if we assume the use unbounded data, channels and number of processes.

Since PROMELA models are intended to specify finite states transition systems, the data types are restricted and the size of channels and the number of active processes is limited to 255. Hence, large firing sequences may not be reproduced by the translation. In general, what we can conclude is that for any firing sequence in the NPN there is an execution path in the PROMELA model corresponding to a prefix of the sequence and vice versa. To reduce the state space, it is important to avoid large bounds for the channels and use data types such as \texttt{unsigned} and \texttt{bit} instead of \texttt{byte} for representing places, colors and labels. Large nets must be analyzed by means of smaller abstract models.

\subsection{Improving the translation}

To conclude we point out that the translation may be improved to reduce the size of model and its efficiency. In particular we highlight the next refinements:

\begin{enumerate}
  \item use priority 1 to create a net token (last row in Table~\ref{tab.act.net.places}) if the initial marking of the element net  can be done in the initialization part of the variable declaration.
  \item the \small\verb"enableTest"\nfont\  in Figure~\ref{figStructNetEl}, line \small\verb"11"\nfont\  should be omitted if the label occurs only once in the element net. It can also be removed if the label occurs several times but in transitions that cannot be enabled at the same time (as transitions $t_3$ and $t_6$ in Figure~\ref{figRPNFact}). In this latter case, the condition should check the identity of the transition instead of its label.
  \item if the labels belonging to $L_h\cup L_v^u$ occur only once in each element net, then use the label as the identity for the transition and get rid of the corresponding message field.
  \item if no net token is transported in the NPN, then the last field of the messages and the \small\verb"if"\nfont-option of Figure~\ref{figStructNetEl}, line \small\verb"19"\nfont\   can be discarded. Furthermore, the \small\verb"if"\nfont\  in Figure~\ref{figStructNetEl}, line \small\verb"21"\nfont\ should be replaced by \small\verb"set_priority(_pid, 1)"\nfont. Since the vertical steps must remove the net tokens involved, then a \small\verb"break"\nfont\   statement should added at the end of the firing any transition with label in $L_v^u$. An  additional \small\verb"if"\nfont-option should be inserted at line \small\verb"19"\nfont\   to cover the situation when the net is removed without synchronization. This option has the form \small\verb"::it==0->break"\nfont.
      The last field of the messages can also be discarded if no net token is consumed. The \small\verb"if"\nfont\  in Figure~\ref{figStructNetEl}, line \small\verb"21"\nfont\ should be replaced as before. This refinement may be applied to individual element nets even when the last field cannot be removed.
\end{enumerate}

We have used some of these hints for translating the element net $F$ of our running example. Its \small\verb"proctype"\nfont\ definition is shown in Figure~\ref{figTransF} (we have defined $\bar{\lambda}=1$). The complete PROMELA model can be found at \small\url{http://www.ime.usp.br/~mirtha/factExImproved.pml}\nfont.

\mfont
\begin{figure}[h]
\begin{multicols}{2}
\begin{minipage}[t]{0.5\textwidth}
\begin{verbatim}
proctype netF(chan pc){
byte p6=1,p8; NetPlace p7; f++;
do:: {
 do:: d_step{ p6 >0 &&
     !pc ?? [eval(_pid),1,3,0]->
      pc ! _pid,1,3,0 }
   :: d_step{ p6 > 0 && p1 > 0 ->
      set_priority(_pid, 6);
      p1--; p6--; rmConf(3);
      pc ! _pid,0,4,1 }
   :: d_step{
      p7.d ?? [_,1,_,0] ->
      set_priority(_pid, 6);
      p7.d ?? nt,1,it,0;
      consNetTok(p7.d, nt);
      p7.d ! nt,1,it,1;
      set_priority(nt, 5);
      p8++; set_priority(_pid, 1) }
\end{verbatim}
\end{minipage}

\begin{minipage}[t]{0.5\textwidth}
\begin{verbatim}
   :: d_step{ p8 > 0 &&
     !pc ?? [eval(_pid),1,6,0] ->
      pc ! _pid,1,6,0 }
  od  }
  unless atomic{
  pc ?? eval(_pid),_, it,1 ->
  if:: it == 4 ->
       nt = run netF(p7.d);
       p7.d ! nt, 255,0,0 ;
    :: it == 3 ->
       p6--; p5++; break
    :: it == 6 ->
       p8--; p5++; break
  fi;
  set_priority(_pid, 1) }
od; set_priority(_pid, 1) }
\end{verbatim}
\end{minipage}
\end{multicols}
\caption{PROMELA translation for the element net $F$ in Figure~\ref{figRPNFact}.}
\label{figTransF}
\end{figure}
\nfont

\section{Investigating behavioral properties with SPIN}
\label{sec:prop}

\begin{figure}
\begin{center}
\includegraphics[scale=0.8]{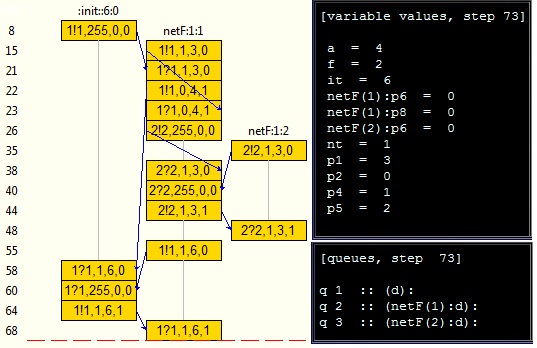}
\end{center}
\vspace{-10pt}
\caption{\texttt{iSPIN} random simulation for the first firing sequence in Example~\ref{figRPNFact}.}
\label{figTraceFact}
\end{figure}

The translation into PROMELA may be help for studying the behavior and verifying properties of NPNs. The simulation facilities provided by SPIN\footnote{The random and interactive simulations provided by the graphical environment iSPIN do not execute embedded C code.}  may be used for testing specific firing sequences from early stages of design (see Figure~\ref{figTraceFact}). But the main advantage of using SPIN is the possibility of detecting sequences violating important properties of a NPN such as termination, reachability and boundedness. In this section we explain how to use SPIN for this purpose.

Termination is a fundamental requirement for many applications. We can test this property with our PROMELA translation using a two-phase approach. In the first phase we should investigate the existence of infinite recursive or unbounded sequences. This can be done using the default verification. To this end, the states of the PROMELA model corresponding to dead markings of the net should be marked as valid \small\texttt{end}\nfont states for SPIN. Such states are those in which every active process is blocked at the any of the loops. If an infinite recursive or unbounded sequence exists then the model has infinite states; but SPIN enforces the finiteness restriction by limiting the number of active processes and channels. Therefore, such infinite sequence will lead to an invalid state that will be reported by the SPIN default verification. Nevertheless, a long terminating sequence going beyond the bounds of the model or SPIN limits, will also produce an invalid state. Using SPIN guided simulation or advanced options (e.g. for increasing the vector size) we may get insights of the real situation. When the search is completed without errors, the verifier may report some unreachable states. One of these states should be the end of \texttt{init} since $SN$ cannot be consumed. Other unreachable states may be due to transitions which never fire or net tokens which are never consumed. This information may be help to improve the NPN definition.

In the second phase we should investigate the infinite cycles in the net. In this case we should use the option for searching for acceptance cycles. A state of a PROMELA model is an acceptance state if any of the active processes is at a statement labeled with an acceptance label. The verifier generated by SPIN will report any infinite run that visit an acceptance state infinitely often. Therefore, for our translation, it is enough to add an acceptance label in front of the loop of \small\texttt{init}\nfont. Note that this will mark the states starting the firing of any transition, and hence the beginning of a step. If the net has a cycle, then there is an acceptance state in the PROMELA model that belongs to cycle in the state space of the model. If the new search is completed and no acceptance cycle is found we can conclude that the NPN is terminating.

\begin{example}
\label{ex:spin-prop1-fact}
Since the NPN of our running example has no cycle, termination can be proved by a default verification of the PROMELA model. SPIN (version 6.2.7) took $0.001s$ for this analysis, running in a notebook Intel Core I3, 2.4GHz, 4Gb RAM. For $p1=39$, the time was $0.058s$ but for $p1 =40$ the verification could not be completed. The same happens if, as in~\cite{Lomazova01}, we remove the places \small\verb"p1"\nfont\ and \small\verb"p5"\nfont, leading to the next infinite recursive sequence (we have omitted the places in the markings).

\vspace{7pt}
$\lessdot 1,\emptyset, 0\gtrdot\mathbf{[t_1\rangle}$

$\lessdot 0, (F,\lessdot 1,\emptyset,0\gtrdot),0\gtrdot \ \mathbf{[t^1_4\rangle}$

$\lessdot 0, (F,\lessdot 0,(F,\lessdot 1,\emptyset,0\gtrdot),0\gtrdot),0\gtrdot\ \mathbf{[t^2_4\rangle}$

$\lessdot 0, (F,\lessdot 0,(F,\lessdot 0,(F,\lessdot 1,\emptyset,0\gtrdot),0\gtrdot),0\gtrdot),0\gtrdot\ \mathbf{[t^3_4\rangle} $

$\lessdot 0, (F,\lessdot 0,(F,\lessdot 0,(F,\lessdot 0,(F,\lessdot 1,\emptyset,0\gtrdot),0\gtrdot),0\gtrdot),0\gtrdot),0\gtrdot \mathbf{[t^4_4\rangle} t\ldots $

\vspace{7pt}

In both cases the same error trail is obtained. However, for the terminating version, increasing the size of state vector (-DVECTORSZ=2048) was enough for completing the verification for $p1=80$ in $0.436s$.
\end{example}


Boundedness can be easily studied with the default verification using assertions on the variables for places. The assertions should be placed after the \small\verb"produceActions"\nfont\ of each transition with an incident arc to the required places. Reachability conditions can be encoded using LTL properties or never claims. To this end, the involved places of $SN$ should be declared as global variables. Conditions involving net tokens should be specified as  a \small\texttt{never}\nfont-claim, in order to access the local places through remote references. However, this is only possible if the total number of net tokens is known in advance. Never-claims may also help to analyze properties of nets with cycles. Since SPIN  provides support just for process-level weak fairness, conditions enforcing the strong fairness should be embedded inside the claim.

\begin{example}
The last statement in Example~\ref{ex:npn-fact} can be proved with SPIN. To this end, we declared all places of $SN$ as global variables. Besides, we included two additional variables: $a$ for saving the initial marking of \small\verb"p1"\nfont\ and $f$ for counting the number of net tokens created. The statement can be specified using the LTL property \small\verb" <>[](p4==1 && p2==0 && len(p3.d)==0 && p1==a-f+1 && p5==f)"\nfont. The analysis took $0.945s$ for  $p1=80$.
\end{example}

\section{Experiments}
\label{sec:performance}

We have run some experiments to assess the effectiveness of SPIN for simulating and analyzing NPNs. In all cases, we were able to fully verify only small instances of the examples. We believe that this is due to the intensive use of channels that our translation demands. PROMELA provides a poor support for channel-based operations such as the non-deterministic choice of an element and the removal (transference or the number) of elements matching a pattern. These operations implemented as predefined statements may help to improve the readability of the simulations and reduce the complexity of the verifications. Furthermore, in our translation, all channels represent multisets; therefore, the state space may include a large number of execution paths representing the same firing sequence.

The above issues are best handled using a C data type instead of a channel for representing the colored places. However, even with a sophisticated representation and functions for dealing with multisets, we may not obtain a significant reduction of the state space, without effective reduction strategies. Unfortunately, process priorities require compilation without the partial order reduction (POR). Therefore, we also designed a version of the translation that attaches the priorities to the response messages instead of the processes. It is simpler (though less intuitive) than the original one: it has a single loop for the element nets and no \small\verb"unless"\nfont\  statement. Besides, all response messages are sent via the global channel, ordered according to its priority. See the  details in Appendix~\ref{app:alt.trans}

\begin{figure}
\begin{center}
\includegraphics[scale=0.4]{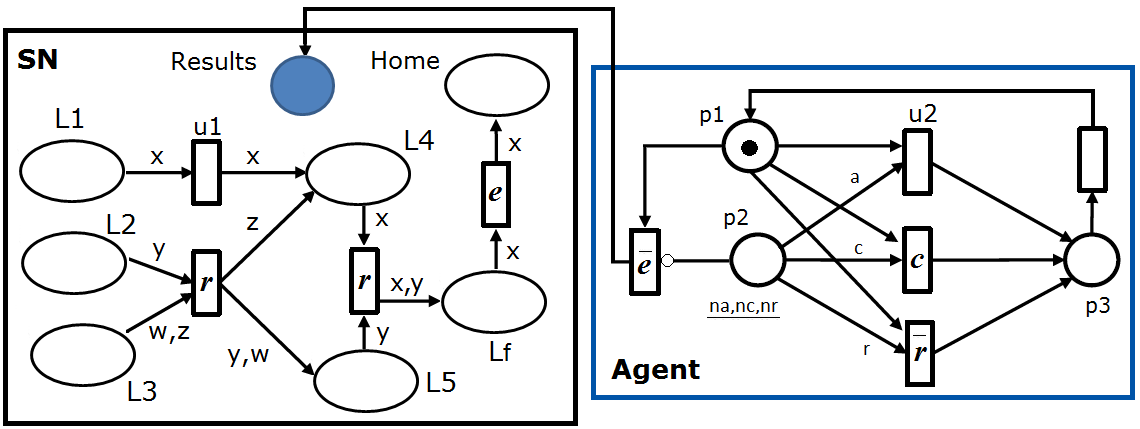}
\end{center}
\vspace{-10pt}
\caption{NPN for a multi-agent based scenario.}
\label{figEx1}
\end{figure}

In the later, we use an example to illustrate SPIN performance using both approaches. The example models an scenario where agents must perform some tasks autonomously or in collaboration. To this end, each agent traverses the environment, collaborating (if possible) with agents at the same location. The environment may also enforces the collaboration by coupling agents situated at different locations. An agent can move back to its source location once it completes all assigned activities. Figure~\ref{figEx1} shows a NPN modeling a simple environment (${\it SN}$) and the agents. The circles represent basic colored places while the ellipses represent net-typed places. The initial marking for an agent net has an uncolored token at $p1$, a number of tokens $a,c,r$ at the colored place $p2$ and $p3$ empty, hereafter denoted as $Agent(na, nc, nr)$. The basic colors  $a,c,r$ represent tasks that are executed autonomously ($a$), in collaboration ($c$) or as a request from the environment ($r$). The  transitions labeled as $c$ and $r$ are used for horizontal and vertical synchronization resp. An agent moves to \emph{Home} by means of transitions labeled as $e$ and $\bar{e}$, for vertical synchronization. The latter transition has an inhibitor arc\footnote{An inhibitor arc tests the absence of tokens at the input place.
It is represented using a line with a circle instead of an arrow on the transition side.} that enables its firing once all tasks have been completed. Its firing adds a token to the shared place $Results$.

The PROMELA model for this NPN\footnote{\small\url{http://www.ime.usp.br/~mirtha/exMASEnv.pml}\nfont} can be used to investigate if from an initial configuration of the environment, all the agents \emph{always} reach the \emph{Home} place. Due to the NPN definition, this property can easily specified as \small\verb" <>(Results==NumberAgents)"\nfont. An example of such initial configuration (hereafter called \emph{sound}) is the marking $I_0^1$ where $L1=Agent(1, 0, 1),L2=Agent(1, 1, 2),$ $L3=\{Agent(1, 1, 2),Agent(2, 1, 2)\}$, the remaining places in ${\it SN}$ are empty and the arity of $c$ is 3. Another sound configuration used in our experiments is e.g. $I_0^2$ that differs from $I_0^1$ in the number of autonomous steps of all net tokens (2). A sound configuration turns up unsound just with a slight modification, e.g. $I_0^1$ with  $ar(c)=2$ (denoted as $I_0^3$) or with another identical agent at $L1$ (denoted as $I_0^4$). The next table shows the results of a safety verification for these markings. The rows with the first cell marked as -P+R correspond to the translation without priorities and using POR. The time for verifying the property is shown in the last column. As the table shows, the verification with priorities scales better than removing this feature and using POR. This is because the effectiveness of the POR technique decreases as the dependencies between the net processes grow.

\vspace{10pt}
\begin{center}\begin{tabular}{|c|c|c|c|c|c|c|}
      \hline
      \ \ Initial\ \  &\ \ States\ \ &\ Time (s)\ \ &\ Memory (Mb)\ \ &\ Extra\ &\ \ Time (s)\ \ \\
      \ \ Marking\ \  &\ \ \ \ &\ \ \ &\ \ \ &\ Options\ &\ \ Property \ \ \\
      \hline
      $I_0^1$  & 2492678  & 25.7 & 1122.878 &  -DMEMLIM=  &  57.1 \\
             &          &      &          &      4096   &    \\
      \hline
      -P+R        & 1643699  & 12.5 & 497.194 & - & 81.4  \\
      \hline
      $I_0^2$ & 9283833  & 95.8 &   352.272 & -DCOLLAPSE &  215 \\
      \hline
      -P+R       & 11362815  & 156 &   414.148   & -DCOLLAPSE &  360 \\
      \hline
      $I_0^3$ & 5829875  & 53.1 &   252.175 & -DCOLLAPSE &   0.002 \\
      \hline
      -P+R      & 7089339  &  92.1 & 287.526 & -DCOLLAPSE & 0.001 \\
      \hline
      $I_0^4$ & 63157720  & 496 & 166.342 & -DBITSTATE & 0.002 \\ 
      \hline
      -P+R        &  105366760  & 1.09e3 & 162.528 & -DBITSTATE & 0.002 \\ 
      \hline
\end{tabular}\end{center}
\vspace{10pt}

The translation with priorities can be improved by dynamically fixing the order in which the processes are executed in a synchronizing step. To this end, we  use the global channel as in the version without priorities. This avoids the interleaving of firings inside the step and may reduce the state space of the model. Besides, it does not affect the correctness of the translation since the  transitions involved in a step do not share input places. The details of the improved translation appear in Appendix~\ref{app:imp.trans}.

\begin{figure}
\begin{center}
\includegraphics[scale=0.35]{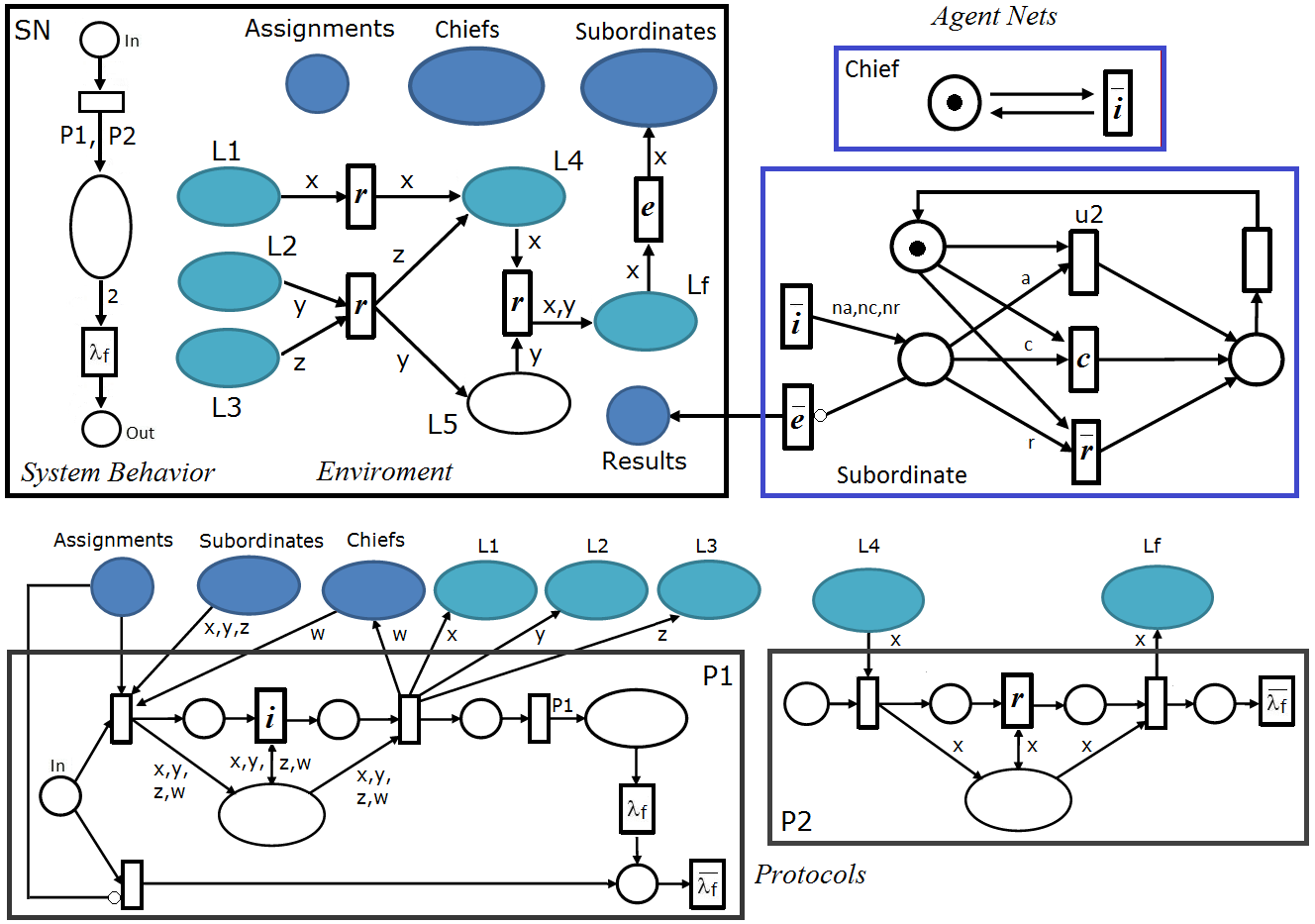}
\end{center}
\vspace{-10pt}
\caption{NPN for a multi-agent based scenario.}
\label{figEx2}
\end{figure}

In the later we use a larger variant of the NPN in Figure~\ref{figEx1} to illustrate the results obtained when comparing the three versions. The net, depicted in Figure~\ref{figEx2}, has four element nets: two types of agent nets and two others for coordination protocols. The system net comprises two subnets for the environment and the system behavior. The shared places are colored in blue. The element net $P1$ is recursive.

The next table summarizes the results obtained for the model. The letters in the first column indicate the translation used: +P-R for the original translation, +P+F-R for improved translation\footnote{\small\url{http://www.ime.usp.br/~mirtha/exMASEnvExt.pml}\nfont} and -P+R (resp. -P-R) for the version without priorities compiled using (resp. without using) POR. We used two initial configurations (C1-sound and C2-unsound) with three agents at $Subordinates$, an agent at $Chiefs$ and a single token at $Assigments$. In all cases, the safety verification was completed using bit state hashing, though with a small hash factor. Note that, the time for completing the search  when using the improved translation, in spite of the fact that the vector size and the states (and hence the memory requirements) are slightly larger.

\vspace{10pt}
\begin{center}\begin{tabular}{|c|c|c|c|c|c|c|c|}
      \hline
      \ \ Translation\ \  &\ \ States\ \ &\ Time\ \ &\ Memory\ \ &\ \ State-\ \ &\ \ Time\ \ &\ \ Time\ \ \\
      \ \  \ \  &\ \  \ \ &\ \ \ &\ \ (Mb) \ &\ vector\ &\ \ C1 \ \ &\ \ C2 \ \ \\
      \hline
      +P-R  & 597287  & 4.06 & 418.099 &  730  &  7.69 & 0.002  \\
      \hline
      -P+R & 596985  & 4 &   413.333 & 722 &   7.75 & 0.002\\
      \hline
      -P-R & 600335  &  4.7 &   415.652 & 722 &   8.87 & 0.002 \\
      \hline
      +P+F-R  & 598004  & 3.91 & 423.163 & 738 & 7.49  & 0.002 \\
      \hline
\end{tabular}\end{center}
\vspace{10pt}

For this example, we use a more complex LTL property for ensuring the agents reach $Subordinates$ and the protocols terminate:
\small
\begin{verbatim}
<>[]( len(L1.d)==0 &&  len(L2.d)==0 &&  len(L3.d)==0 &&  len(L4.d)==0 &&
      len(L5.d)==0 &&  len(Lf.d)==0  &&  Assig==0 &&  pwOut==1 )
\end{verbatim}
\nfont
\noindent Note that the number of tokens at a place may not be equal to the length of the channel since the channel may also contain several request messages. A simpler property is obtained if we use an additional global integer variable for counting the number of net tokens at $Subordinates$. This variable should be updated each time the marking of the place is modified by the firing of a transition.

The results for C1 and C2 are shown in the last two rows of the above table.  We also used more demanding sound and unsound configurations obtaining similar results. In all cases, advanced options (-DVECTORSZ, -DBITSTATE, -DCOLLAPSE) were required for completing the safety verification and proving valid properties. But in most of the cases, the basic search was enough for finding a counterexample to invalid properties.

\section{Conclusions and Future Work}
\label{sec:conclusion}

This paper presented a general translation from NPNs (with arbitrary levels and recursion) into PROMELA. The resulting models can be used to analyze termination, boundedness and LTL properties of the underlying nets. The current translation is best suited for detecting firing sequences violating the properties. But we are currently working on a refined version that may help to fully verify larger models. In addition, we plan to implement a plug-in for integrating SPIN with a visualization tool such as Renew. We will use the ideas behind the translation as guidance for using other model checkers such as JPF2 and Maude. We are interested on comparing the performance of these tools using models in the areas of cross-organizational workflows and multi-agent systems.  Reduction strategies that can be effective for model checking these nets will also be subject of future work.

\bibliographystyle{plain}
\bibliography{MBibliography}

\newpage

\appendix

\section{Auxiliary functions}
\label{app:aux.code}

\mfont
\begin{figure}[h!]
\begin{multicols}{2}
\begin{minipage}[t]{0.5\textwidth}
\begin{verbatim}
byte nt,lt,it; bit rm;
NetPlace cha;
byte v0,v1,v2;

/* ch ?* f0,f1,f2,0,0 */
inline recMsg(ch,f0,f1,f2){
 do:: ch ?? [f0,f1,f2,0,0] ->
        ch ?? f0,f1,f2,0,0;
        cha.d ! f0,f1,f2,0,0;
   :: else -> break
 od;
 cha.d ? f0,f1,f2,0,0;
 do:: cha.d ?? [_,_,_,_,_]->
      if :: cha.d ? v0,v1,v2,0,0;
              ch  ! v0,v1,v2,0,0;
         ::   ch  ! f0,f1,f2,0,0;
            cha.d ? f0,f1,f2,0,0;
      fi
   :: else -> break
  od; skip }
\end{verbatim}
\end{minipage}

\begin{minipage}[t]{0.5\textwidth}
\begin{verbatim}
inline consNetTok(ch, p){
  do:: ch ?? [eval(p),_,_,0,0] ->
       ch ?? eval(p),_,_,0,0;
    :: else -> break
  od; skip }

inline consNetsAtPlace(ch){
 do:: ch ?? [_,255,0,0,0] ->
      ch ?? nt,255,0,0,0;
      consNetTok(ch, nt);
      ch ! nt,0,0,1,1;
      set_priority(nt, 3);
   :: else -> break
 od; skip }

inline transpNetTok(ch, och, p){
  do:: ch ?? [eval(p),_,_,_,_] ->
       ch ?? eval(p),v1,v2,0,0;
       och ! p,v1,v2,0,0;
    :: else -> break
  od; skip }
\end{verbatim}
\end{minipage}
\end{multicols}

\begin{verbatim}


c_code{  typedef struct QNP {
             uchar Qlen; /* q_size */
             uchar _t;   /* q_type */
             struct {
                 uchar fld0, fld1, fld2;
                 unsigned fld3 : 1, fld4 : 1;
             } contents[MaxMsg]; } QNP;

         int numMsg(uchar *z, int lab){
             int n = ((Q0 *)z)->Qlen, c = 0;
             for (int k = 0; k<n; k++)
               if ( ( ((QNP *)z)->contents[k].fld1 == lab )  &&
                    ( ((QNP *)z)->contents[k].fld3 == 0   ) )   c++;
             return c; }
};

/* A call to numMsg has the form numMsg(qptr(PProcName->c - 1), v)
   where c is a channel and v is an integer expression. The prefix
   "PProcName->" (e.g. Pinit->) is used to refer a local variable
   inside a c_expr. For a global variable the prefix is "now." . */
\end{verbatim}

\vspace{-10pt}
\caption{Auxiliary variables, inline definitions and  C functions.}
\label{app1}
\end{figure}
\nfont

\newpage

\section{Translation without process priorities}
\label{app:alt.trans}

See Figure~\ref{figStructNPNwoP} and Tables~\ref{tab.opt.el.net.wo.p} and~\ref{tab.act.net.places.wo.p} for details. The model corresponding to this translation for the NPN in Figure~\ref{figEx2} can be found at \small\url{http://www.ime.usp.br/~mirtha/exMASEnvExtNP.pml}\nfont.

\mfont
\begin{figure}[h!]
\begin{center}
\begin{verbatim}
 1   typedef NetPlace { chan d = [MaxMsg] of {byte,byte,byte} }
 2   chan gbChan = [MaxMsg] of {byte, byte, byte, chan, bit};
 3
 4    /* Auxiliary Code, Shared Places, Element Nets */
 5
 6  proctype EN_i(){
 7  chan ppChan;  /* Non-Shared Places and Arc Variables */
 8  atomic{ gbChan ? 2,eval(_pid),255,ppChan,0;
 9          /* Initial Marking */ }
10  do :: atomic{/* autonomous firing          - Op1 */}
11     :: d_step{/* synchronization request    - Op2 */}
12     :: d_step{/* horizontal synchronization - Op3 */}
13     :: ...
14     :: atomic{ gbChan ? _,eval(_pid),lt,ppChan,rm  ->
15           if :: lt==L(t) && enableTest_t ->    /* sync step */
16                        consumeActions_t;
17                        produceActions_t; rmConf_t;
18              :: ...
19              :: lt==255 -> skip
20           fi;
22           if:: rm -> break :: else fi }
22  od;
23  d_step{ consNetsAtPlace(np1); ... }
24  set_priority(_pid, 1) }
25  }
26
27  init(){
28  /* Non-Shared Places and Arc Variables */
29  atomic{ /* Initial Marking */ }
30  do :: atomic{ empty(gbChan) && enableTest_t ->
31            consumeActions_t; produceActions_t; }
32     :: ...
33  od }
\end{verbatim}
\end{center}
\vspace{-5pt}
\caption{General structure of a PROMELA model for a NPN.}
\label{figStructNPNwoP}
\end{figure}
\nfont

\small
\begin{table}[h!]
\begin{center}\begin{tabular}{|l|l|}
      \hline
      \verb" Op1 "  & \verb" empty(gbChan) && enableTest_t -> "  \\
                    & \verb"       consumeActions_t;"  \\
                    & \verb"       produceActions_t; rmConf_t;"  \\
      \hline
      \verb" Op2 "  & \verb" empty(gbChan) && enableTest_t && ! ppChan??[eval(_pid),L(t),_] ->  "\\
                    & \verb"       ppChan!_pid,L(t),t" \\
      \hline
      \verb" Op3 "  & \verb" empty(gbChan) && ppChan??[eval(_pid),Lh,_] && "  \\
                    & \verb" c_expr { numMsg(qptr(PEN_i->ppChan-1), Lh) >= ar(Lh) } ->  "  \\
                    & \verb"       /* repeat the next code ar(Lh) times */ "  \\
                    & \verb"       ppChan ?* nt,Lh,it; "  \\
                    & \verb"       gbChan !! 4,nt,Lh,ppChan,0; "  \\
      \hline
\end{tabular}\end{center}
\caption{Different options in the loop of an element net.}
\label{tab.opt.el.net.wo.p}
\end{table}
\normalsize

\small
\begin{table}[h!]
\begin{center}\begin{tabular}{|l|l|}
      \hline
      \verb" consumeAction(_)" & \verb" np.d ?* nt,-Lvl(t),it; "    \\
                               & \verb" consNetTok(np.d,nt);"          \\
                               & \verb" if :: Lvl(t)==0 -> " \\
                               & \verb"               gbChan !! 3,nt,-Lvl(t),np.d,1; " \\
                               & \verb"    :: else -> gbChan !! 5,nt,-Lvl(t),np.d,1; "  \\
                               & \verb" fi"                   \\
      \hline
      \verb" transportAction(x) "        & \verb" np.d ?* nt,-Lvl(t),it;"\\
      \verb"   np == opn && Lvl(t)>0  "  & \verb" gbChan !! 5,nt,-Lvl(t),np.d,0;"   \\
      \hline
      \verb" transportAction(x)" & \verb" np.d ?* nt,-Lvl(t),it;"\\
      \verb"   np != opn "         & \verb" transpNetTok(np.d,onp.d,nt);" \\
                                 & \verb" if :: Lvl(t)==0 -> " \\
                                 & \verb"               np.d ! nt,255,0;  " \\
                                 & \verb"               gbChan !! 3,nt,-Lvl(t),onp.d,0; " \\
                                 & \verb"    :: else -> gbChan !! 5,nt,-Lvl(t),onp.d,0; " \\
                                 & \verb" fi"                   \\
      \hline
      \verb" produceAction(EN_i)" & \verb" nt = run ElementNet_i();"\\
                                  & \verb" np.d ! nt,255,0; "\\
                                  & \verb" gbChan !! 2,nt,255,np.d,0; "\\
      \hline
\end{tabular}\end{center}
\vspace{-2pt}
\caption{Consume, produce and transport actions for net-typed arc labels.}
\label{tab.act.net.places.wo.p}
\end{table}
\normalsize

\newpage
\section{Translation with priorities and fixing the synchronization ordering}
\label{app:imp.trans}

See Figure~\ref{figStructNPNimpP} and Tables~\ref{tab.opt.el.net.imp.p} and~\ref{tab.act.net.places.imp.p} for details. The model corresponding to this translation for the NPN in Figure~\ref{figEx2} can be found at \small\url{http://www.ime.usp.br/~mirtha/exMASEnvExtImp.pml}\nfont

\mfont
\begin{figure}[h!]
\begin{center}
\begin{verbatim}
 1   typedef NetPlace { chan d = [MaxMsg] of {byte,byte,byte} }
 2   chan gbChan = [MaxMsg] of {byte, byte, byte, chan, bit};
 3
 4    /* Auxiliary Code, Shared Places, Element Nets */
 5
 6  proctype EN_i(){
 7  chan ppChan;  /* Non-Shared Places and Arc Variables */
 8  atomic{ gbChan ? 6-2,eval(_pid),255,ppChan,0;
 9          /* Initial Marking */;
10          set_priority(_pid, 1) }
11  do :: atomic{/* autonomous firing          - Op1 */}
12     :: d_step{/* synchronization request    - Op2 */}
13     :: d_step{/* horizontal synchronization - Op3 */}
14     :: ...
15     :: atomic{ gbChan ? _,eval(_pid),lt,ppChan,rm  ->
16           if :: lt==L(t) && enableTest_t ->    /* sync step */
17                        consumeActions_t;
18                        produceActions_t; rmConf_t;
19              :: ...
20              :: lt==255 -> skip
21           fi;
22           if:: rm -> break :: else fi;
23           set_priority(_pid, 1) }
24  od;
25  d_step{ consNetsAtPlace(np1); ... }
26  set_priority(_pid, 1) }
27  }
28
29  init(){
30  /* Non-Shared Places and Arc Variables */
31  atomic{ set_priority(0,2); /* Initial Marking */; set_priority(0,1) }
32  do :: atomic{ empty(gbChan) && enableTest_t ->
33            set_priority(_pid, 6);
34            consumeActions_t; produceActions_t;
35            set_priority(_pid, 1)
36        }
37     :: ...
38  od }
\end{verbatim}
\end{center}
\vspace{-5pt}
\caption{General structure of a PROMELA model for a NPN.}
\label{figStructNPNimpP}
\end{figure}
\nfont

\small
\begin{table}[h!]
\begin{center}\begin{tabular}{|l|l|}
      \hline
      \verb" Op1 "  & \verb" empty(gbChan) && enableTest_t -> "  \\
                    & \verb"       set_priority(_pid, 6);"  \\
                    & \verb"       consumeActions_t;"  \\
                    & \verb"       produceActions_t; rmConf_t;"  \\
                    & \verb"       set_priority(_pid, 1);"  \\
      \hline
      \verb" Op2 "  & \verb" empty(gbChan) && enableTest_t && ! ppChan??[eval(_pid),L(t),_] ->  "\\
                    & \verb"       ppChan!_pid,L(t),t" \\
      \hline
      \verb" Op3 "  & \verb" empty(gbChan) && ppChan??[eval(_pid),Lh,_] && "  \\
                    & \verb" c_expr { numMsg(qptr(PEN_i->ppChan-1), Lh) >= ar(Lh) } ->  "  \\
                    & \verb"       set_priority(_pid, 6);"  \\
                    & \verb"       ppChan ?? _pid,Lh,it; "  \\
                    & \verb"       gbChan !! 6-6,nt,Lh,ppChan,0; "  \\
                    & \verb"       /* repeat the next code ar(Lh)-1 times */ "  \\
                    & \verb"       ppChan ?* nt,Lh,it; "  \\
                    & \verb"       gbChan !! 6-4,nt,Lh,ppChan,0; "  \\
                    & \verb"       set_priority(nt, 4);"  \\
      \hline
\end{tabular}\end{center}
\caption{Different options in the loop of an element net.}
\label{tab.opt.el.net.imp.p}
\end{table}
\normalsize

\small
\begin{table}[h!]
\begin{center}\begin{tabular}{|l|l|}
      \hline
      \verb" consumeAction(_)" & \verb" np.d ?* nt,-Lvl(t),it; "    \\
                               & \verb" consNetTok(np.d,nt);"          \\
                               & \verb" if :: Lvl(t)==0 -> " \\
                               & \verb"               gbChan !! 6-3,nt,-Lvl(t),np.d,1; " \\
                               & \verb"               set_priority(nt, 3);" \\
                               & \verb"    :: else -> gbChan !! 6-5,nt,-Lvl(t),np.d,1; "  \\
                               & \verb"               set_priority(nt, 5);" \\
                               & \verb" fi" \\
      \hline
      \verb" transportAction(x) "        & \verb" np.d ?* nt,-Lvl(t),it;"\\
      \verb"   np == opn && Lvl(t)>0  "  & \verb" gbChan !! 6-5,nt,-Lvl(t),np.d,0;"   \\
                                         & \verb" set_priority(nt, 5);" \\
      \hline
      \verb" transportAction(x)" & \verb" np.d ?* nt,-Lvl(t),it;"\\
      \verb"   np != opn "         & \verb" transpNetTok(np.d,onp.d,nt);" \\
                                 & \verb" if :: Lvl(t)==0 -> " \\
                                 & \verb"               np.d ! nt,255,0;  " \\
                                 & \verb"               gbChan !! 6-3,nt,-Lvl(t),onp.d,0; "   \\
                                 & \verb"               set_priority(nt, 3);" \\
                                 & \verb"    :: else -> gbChan !! 6-5,nt,-Lvl(t),onp.d,0; "   \\
                                 & \verb"               set_priority(nt, 5);" \\
                                 & \verb" fi"                   \\
      \hline
      \verb" produceAction(EN_i)" & \verb" nt = run ElementNet_i() priority 2;"\\
                                  & \verb" np.d ! nt,255,0; "\\
                                  & \verb" gbChan !! 6-2,nt,255,np.d,0; "\\
      \hline
\end{tabular}\end{center}
\vspace{-2pt}
\caption{Consume, produce and transport actions for net-typed arc labels.}
\label{tab.act.net.places.imp.p}
\end{table}
\normalsize

\end{document}